# Investigating the Impact of Global Positioning System Evidence

Kiyoshi J Berman
University of Glasgow
kiyoshiberman@yahoo.com

William Bradley Glisson
University of South Alabama
bglisson@southalabama.edu

L. Milton Glisson
Ret. N.C. A&T State University
lglisson@triad.rr.com

## Abstract

*The continued amalgamation of Global Positioning Systems (GPS) into everyday activities stimulates the idea that these devices will increasingly contribute evidential importance in digital forensics cases. This study investigates the extent to which GPS devices are being used in criminal and civil court cases in the United Kingdom through the inspection of Lexis Nexis, Westlaw, and the British and Irish Legal Information Institute (BAILII) legal databases. The research identified 83 cases which involved GPS evidence from within the United Kingdom and Europe for the time period from 01 June 1993 to 01 June 2013. The initial empirical analysis indicates that GPS evidence in court cases is rising over time and the majority of those court cases are criminal cases.*

## 1. Introduction

The increasing pervasive integration of Global Positioning Systems (GPS) into today's globally integrated atmosphere encourages an environment that is dependent on accurate information for guidance. This type of information is valuable to a variety of modes of transportation that include air, maritime and land. Not only do these devices help with navigation but they also, potentially, help with defense, search and rescue operations and meteorology.

ABIreserach predicts that personal tracking applications and devices will grow with a compounded average growth rate of 40%, enabling both markets to hit the $1 billion sales mark in 2017 [1]. The news release by ABIreserach goes on to forecast that this will translate into 2.5 million hardware units in 2017 with a significant focus on health care, elderly and solitary worker markets. The rapid growth of these devices can be attributed, in part, to their integration with other technologies. GPS technology is being embedded into everything from smartphones, to watches, to walking sticks, to shoes, to dog collars [15]. The sustained amalgamation of GPS into a range of technologies and devices raises complex ethical and legal debates regarding liability, and privacy with specific concerns focusing on employer monitoring [17].

Liability issues surface when police and private individuals make incorrect decisions based on GPS data. In the United Kingdom, police mistakenly forced their way through the wrong door based on iPhone's GPS software directions [30]. In Ohio, a bank repossessed the wrong house and the contents based on incorrect GPS data [2, 4].

Privacy issues are deliberated when GPS devices are offered as a cost-effective solution for monitoring elderly individuals [27], tracking children to keep them safe [25] and observing employees in the workplace [22]. The issue has escalated as high as the Supreme Court. In the United States v Jones case, the court ruled that a valid search warrant is required to monitor a vehicle's movements for four weeks [12].

In addition to legal discussions, the Wall Street Journal published an article on the use of a mobile phone's internal GPS chip for stalking, along with instances where it led to assault and murder [24]. A transportation threat has been demonstrated through the successful injection of spoofing signals into the GPS navigation of a super-yacht [21]. The rapid evolution of GPS devices into compact components used to improve transportation and location services can also be used to initiate and sustain illegal activities that impact individuals, businesses and, potentially, national security.

Based on the impact of this information, it is hypothesized that the use of GPS evidence, in court proceedings, has increased during the past two decades and has, increasingly, played a critical role in court rulings. The hypotheses raised several subsidiary research questions that needed to be explored in order to address the hypotheses:

1. Is it possible to investigate the extent to which the information about GPS evidence is explained in the court cases and determine if the criticality of the data can be quantified?
2. Can correlations between the type of crime and the transportation mode of GPS evidence be identified?
3. Are trends identifiable in the GPS evidence presented in court cases during the past two decades?

This research identified 83 cases which involved GPS evidence within the UK and Europe for the time period from 01 June 1993 to 01 June 2013.

## 2. Relevant work

Recent research indicates that mobile device evidence is having an increased impact in United Kingdom court cases [16]. While GPS technology is typically an option for mobile devices, it is increasingly being integrated into a variety of other aspects of technology. The capabilities, limitations and overall impact of GPS devices are being investigated in relation to transportation [18, 26, 29] and surveillance [3, 10, 11, 31]. In addition, researchers are investigating variables that impact, along with solutions to improve, the technical accuracy of GPS devices [5, 9, 28].

While the traditional modes of transportation, e.g., Air, Water, Rail, Motor and Pipe have been well defined for several years [20], GPS technologies are being researched from the logistics and technological perspectives. From the logistics viewpoint, research is attempting to establish location requirements for logistics companies servicing motor, aviation and maritime transportation modes [18]. From a technical standpoint, Ta, et. al.,[29] presents a wireless ad hoc network that is designed to provide location services to small fishing boats, while improving the Sea-to-land radio links connecting boats to inland stations. The proposed system consists of GPS receivers with a wireless network.

Shih, et. al., [26] investigates optimal control patterns for ship maneuvers to avoid collisions. The aim of this research is to find a formulation that would allow optimal maneuvering in any sea condition. The authors present a model on which future studies can be based. While this research doesn't provide a solution for ship maneuvering, it does highlight factors that need to be considered and proposes a "nonlinear unified state-space model" as a strategy to approach the problem [26].

With increasing popularity, decreasing size and cost of GPS enabled devices, it has become easier to track people. Fraser, et. al. [10] examines data from studies to identify the scope of electronic monitoring and stalking. The analysis revealed that out of those who use electronic means to stalk people, 10% use GPS location tracking to locate the victim.

From the law enforcement perspective, GPS technology enables the police to cheaply carry out surveillance and arrest suspects using location based evidence. A range of criminal investigations potentially utilize GPS information [11]. These cases can include murder, drugs, robbery, public corruption and probation violations investigations. However, Ganz [11] acknowledges that there are controversies over the use of surveillance without warrants.

A study conducted by Freeman and Armstrong [3] examines the effectiveness of GPS based monitoring of sex offenders against children kept under community supervision. The data for this study was obtained from Maricopa County, Arizona. The study observed alerts that were set off during a two year period from active and passive GPS devices placed on offenders. The study's results indicate that the loss of satellite signals have triggered false alarms increasing the work load of the officers. The authors indicate that implementing effective policies and practices is increasingly becoming problematic due to a lack of empirical GPS research focusing on implementation.

The use of GPS technology in legal environments forces an examination of the technology from an accuracy perspective. Stergiadou, et. al., [28] discusses factors that directly affect the accuracy of GPS data like atmospheric and multipath interference and receiver clock errors. Atmospheric effects include any atmospheric change that can affect the speed of radio signals. A multipath effect occurs when signals bounce back as a result of hitting any kind of hard object, such as hills or buildings, on the way to the receiver. A receiver-clock error occurs when a receiver does not have a clock that matches the accuracy of the satellite clock.

Drawil, et. al. [9] proposes a methodology to evaluate location estimation accuracy to improve the decision-making capabilities of applications. They incorporate knowledge about the environment into their solution. Costa developed a simulation model that takes into account satellite orbit, elevation maps, buildings, vegetation data, signal fade and multi-channel GPS receivers in order to gage the manipulation of these parameters on performance [5]. Additional errors can be introduced by the use of maps from different online sources which collect data independently and sometimes are not properly updated after traffic configuration changes or new construction is completed [6].

Iqball and Lim [31] argue that law enforcement has not examined the possibilities of GPS data alteration by a person or technical attacks, such as spoofing. The storage design of these devices is open to exploitation. The devices usually do not encrypt tracks, routes and waypoints in the memory allowing it to be easily altered using compatible software tools. The authors indicate that, to date, there are no techniques to validate that the data stored in the device was indeed generated by the device itself; this hinders the non-repudiation quality of evidence. Spoofing attacks are also possible but would be targeting a particular individual rather than a large area. This leaves the possibility for an individual to edit data on a GPS device as a means of implanting evidence against someone. Iqball and Lim [31] also demonstrate the possibility of altering GPS device data by using commercially available GPS devices. This research does not discuss the type of GPS devices used. However, the demonstration includes data alteration using the GPS device software and replacing the flash memory of one GPS device with one of an identical device. They did perform a spoofing attack to demonstrate the vulnerabilities of these GPS devices. Their study is aimed at highlighting the drastic effects which deliberate alteration of GPS data can have on both commercial and legal uses of GPS devices.

There is ample research investigating specific aspects and limitations of GPS technologies along with potential solutions for improvements. However, minimal substantive academic research is being conducted to investigate the impact that GPS devices are having in court room proceedings or to identify trends in conjunction with various modes of transportation.

## 3. Methodology

This research is an historical exploratory case study of court case reports [19]. The study used three major UK legal databases: Westlaw [32], Lexis Nexis [14], and the British and Irish Legal Information Institute (BAILII) [7]. The case study investigates GPS involvement in court cases. The legal libraries were chosen based on availability. The timeline of this research covers the last two decades, from June 1, 1993 to June 1, 2013.

The research was conducted in three stages. In the first stage, the background of Global Positioning Systems was studied to identify 21 keywords for the search process. This included acronyms, new and old synonyms and the most popular brand names of GPS systems. Cases were then identified by using the search terms. In the second stage, a research database was created and data from the identified cases, in the first stage, were recorded in the database. The collected data was then analyzed and organized.

### 3.1 Research scope

This paper focuses on Global Positioning Systems that have been of evidentiary value in court cases. The decision to limit this research to three databases was a pragmatic decision based on resource availability. Therefore, cases not recorded in these databases during the search period or reported in any other database are considered to be out-of-scope for this research.

Also, the legal jurisdictions were limited to the United Kingdom (England, Scotland, Northern Ireland and Wales) and Europe. A further constraint on this paper is that the analysis of the GPS evidence is limited to the information given within the case documentation. For the purposes of this research, the GPS devices were considered on a case association. If a case had multiple GPS devices, the overall impact of the devices was considered for this study, not each individual device. It should also be noted that the data collection for 1993 and 2013 are not complete calendar years.

### 3.2 Research process

The acquisition, the analysis process and the storage of individual court cases in this research was obtained through the following process:
1. Login to the online legal libraries, Lexis and Westlaw, using the University subscription. BAILII does not require a login as it is an open source resource.
2. Once logged in, the 'Case' section of the library was selected. This section contains information on reported cases.
3. In the 'Case' section, all cases within the dates 01 June 1993 and 01 June 2013 were searched individually using the terms provided in Table 1 – Keywords. The terms were derived from journal articles in addition to internet searches to find related systems and the identification of major brand names like Garmin, TomTom and Megallan. It should be noted that the database searches started with LexisNexis, then BAILII, and then Westlaw. A total of twenty-one searches were conducted in each legal database.

4. Once a search had been executed, cases from UK or Europe were individually opened and examined to determine case relevance:
   - If the case involved GPS based evidence, the case was recorded in the database Case table.
   - If the case involved GPS devices, which were not presented as evidence, it was recorded in the IrrelevantCases table. This includes instances where GPS devices were stolen or mentioned in a case regarding patents or taxes. If a GPS device had no evidentiary value, or was not seized, the GPS device memory was not interrogated nor considered to have contained incriminating evidence; it was allocated to this category.
   - If the case was previously recorded or it had no Global Positioning System involved in it at all, then the case was dismissed. It should be noted that the databases were consistently processed in the exact same order.

**Table 1. Keywords**

| GPS |
|---|
| GNSS |
| NAVSTAR |
| GLONASS |
| Galileo |
| Tomtom |
| Garmin |
| Magellan |
| Satnav |
| "global positioning system" |
| "global navigation satellite system" |
| "global navigation" |
| "global positioning satellite" |
| "satellite navigator" |
| "satellite tracking device" |
| "satellite navigation" |
| "navigational system" |
| "radio navigation system" |
| "sat nav" |
| "in-dash navigation " |
| "tracking chip" |

5. If the case was relevant, a classification was assigned based upon the characteristics of how the evidence was used in each case.
   - High: This designation indicates that the GPS evidence in the case was vital to the outcome. A case example in this category would be an arrest based on evidence collected through GPS surveillance or for using a type of GPS device that is illegal in that particular region. The terms used to identify high weight cases included 'highly accurate' and 'mostly depended on'.
   - Medium: This designation indicates that GPS contributed towards the case outcome. In this instance, key aspects such as location, speed or time is determined using GPS to support a main piece of evidence. A case example in this category would be where GPS evidence leads to the location of evidence like a marijuana field. The evidence in these cases is part of a larger compilation and contained terms indicating medium impact such as 'reasonably accurate', 'revealed' and 'no controversial evidence' when describing the evidence.

- Low: This designation indicates that GPS evidence was present or that GPS device(s) related to the defendant were seized for investigation. This would also include GPS evidence which was presented in court and was not questioned further or simply accepted. In this instance, there was a lack of terminology when describing the evidence and no active terms toward the evidence reliance was available.
6. If the case was relevant, a classification was assigned based upon the indicated admissibility of the evidence. Evidence was regarded as *Admitted* in cases where it was clearly stated as accepted/ admitted or if the case clearly states that the GPS evidence affected the outcome of the case. Evidence was regarded as *Dismissed,* if it was clearly rejected or not accepted. However, in cases where there was insufficient information or clarity regarding the admission it was recorded as *Not Available*.
7. The database and all of the relevant material were backed-up to an online drive.

## 4. Results and analysis

The search results identified 281 cases that contained the keywords of which 83 were identified as relevant and assigned a rating of High, Medium or Low. A total of 64 cases were found from Lexis, 8 from Westlaw and 11 from BAILII. It is interesting to note that not all of the keywords provided in Table 1 - Keywords provided relevant results. GPS returned the highest number of cases, 36, while keywords such as 'in-dash navigation' or 'tracking chip' did not appear at all.

Out of the 83 relevant cases identified, 76 cases or 91.6%, were from the United Kingdom. While seven cases or 8.4% were from Europe. Moreover, 55 cases were criminal cases while 28 cases were civil cases. There were 51 appeal cases, of which 49 were criminal and 2 were civil cases. There were 32 initial cases, of which 6 were criminal and 26 were civil cases. For the completeness of the study, there were 198 cases in which GPS devices were mentioned but had no evidentiary value.

### 4.1 Transportation mode analysis

The types of devices involved in the cases were categorized into three modes of transportation, i.e., air, water and land. The categorization is based on where the device was used. The results are displayed in Figure 1 – Mode of transportation. Land accounted for the largest number of cases which was 58. Land cases consisted of 32 automotive cases, 22 cases involved undefined GPS devices, 1 case involved an integrated device, and 3 cases involved GPS tags.

Water accounted for 23 cases and Air for 2 cases. Some of the cases involved the use of more than one device but they were all in the same transportation category. Therefore, the mode of transportation was counted for the number of cases rather than for each individual device. It is interesting to note that only three brand names were mentioned for all cases, these were TomTom, Trimble R8 GNSS and Topcon Hiper Pro.

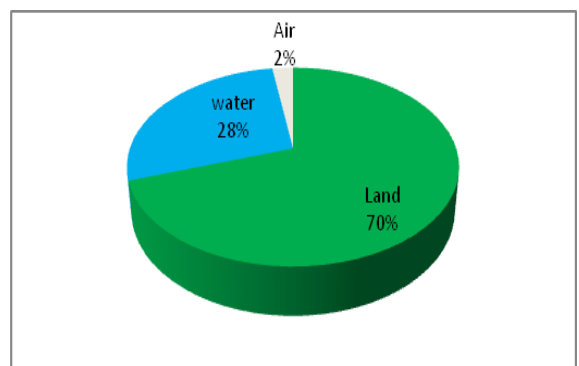

**Figure 1. Mode of transportation**

While Land accounts for the largest number of cases, Figure 2, Transportation mode by year, indicates that the earliest GPS evidence was from the Water mode. There were 15 water cases from 1993 to 2004 prior to cases pertaining to land or air. The water cases involved, mainly, collision disputes between commercial ships and vessels. The number of water cases in this dataset declined after 2005.

The data indicates that there is no GPS evidence involving devices used on land until 2005. However, there is a dramatic rise in the data with 16 cases in 2010, 11 cases in 2011 and 14 cases in 2012. Legal disputes involving GPS devices used in Air-based transportation systems appear in the data starting in 2010. The number of air transportation disputes is still relatively small with 1 case in 2010, no cases in 2011 or 2012 and 1 case in 2013.

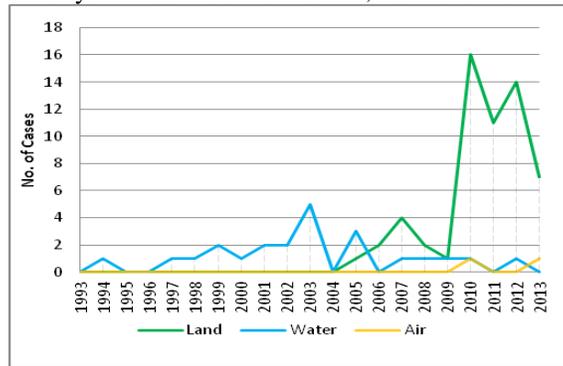

**Figure 2. Transportation mode by year**

Figure 3 – Mode of devices in Criminal Cases and Figure 4 – Mode of devices in Civil Cases presents GPS evidence which appeared in cases according to the mode of transportation.

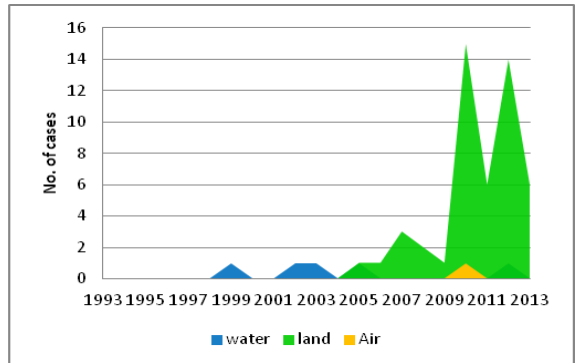

**Figure 3. Mode of devices in criminal cases**

The criminal case data appears to follow a similar pattern of progression as revealed in the transportation mode year analysis. The criminal cases start in water transportation with 3 cases between 1997 to 2003. In 2005, land cases appear with 1 case. Criminal land cases spike in 2010 with 15 cases and in 2012 with 14 cases in correspondence with the data from the transportation mode year analysis.

The dataset indicates that civil cases, as illustrated in Figure 4 – Mode of devices in Civil Cases, started slightly earlier than criminal cases with 2 water transportation cases between 1993 and 1997. While land cases started in 2005, they did not spike until 2011 which recorded 5 cases. Air cases did not appear in the dataset until 2013 and only account for a single case. It is interesting to note that the data indicates that devices used on land were found in most of the criminal litigation with 49 cases. While devices used in the water mode are found in most of the civil litigation with 18 cases.

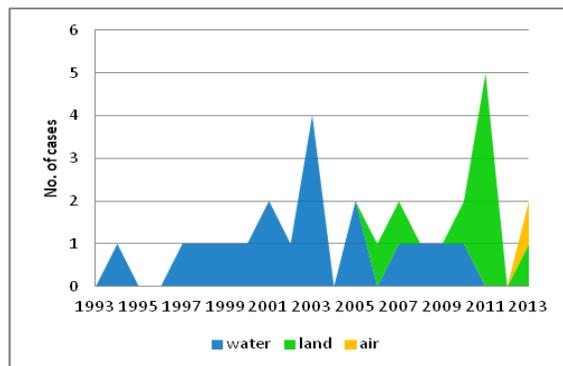

**Figure 4. Mode of devices in civil cases**

## 4.2 Case characteristics

Out of the 83 cases, experts were mentioned in 20 criminal cases and 17 civil cases. In a criminal and a civil case, 2 experts were mentioned in each case. Therefore, a total of 39 experts were identified in the 37 cases. This translates into experts being mentioned in 37 out of 83 or 45% of the cases and not mentioned in 55% of the cases. However, only 7 names were mentioned. Out of the expert titles that were recorded, (forensic analyst, survey expert and ship captain), only one was a digital forensics expert. Typically, the information given about the experts such as names and contact information were not sufficient to carry out further analysis.

It should be noted that, the information pertaining to experts and the technical details of GPS evidence was limited in every case report. There were variances in the presentation of evidence in the case reports. Both issues made it difficult to identity trends in these areas. One explanation for the lack of technical data is that most of the technical details reside in expert reports which were not included in the case reports.

## 4.3 Recorded case types

Overall, there were 83 cases of which 55 cases could be categorized into 19 different criminal classifications and 28 cases could be categorized into 11 different civil classifications. The most recorded crimes were drug related offences as reported in 23 cases which were, predominately, from England and Scotland jurisdictions. There were 15 cases of ship collisions which were from England jurisdiction and prosecuted as civil cases. Table 2 shows the variation of the number of cases according to the types of criminal classifications and Table 3 provides this information for civil classifications.

**Table 2. Criminal Classification**

| Crime Type | Number |
|---|---|
| Use and/or selling of drugs | 16 |
| Importing prohibited drugs | 7 |
| Murder | 4 |
| Murder & Drug dealing | 1 |
| Murder & Conspiracy to rob | 1 |
| Conspiracy to effect illegal entries into the United Kingdom | 2 |
| Conspiracy to rob | 3 |
| Conspiracy to disguise criminal property | 1 |
| Theft | 3 |
| Burglary | 2 |
| Terrorism Acts/ terrorism Prevention | 4 |
| Speeding, dangerous driving | 1 |
| dangerous driving & vehicle collision | 2 |
| Rape | 2 |
| Robbery | 2 |
| Possession of prohibited firearm | 1 |
| Trespassing | 1 |
| Shipping and navigation – failure of satellite tracker | 1 |
| Use of jamming device/transponder | 1 |

**Table 3. Civil Classification**

| Case type | Number |
|---|---|
| Shipping and navigation – Collision | 15 |
| Shipping and navigation - vessel capacity | 1 |
| Shipping and navigation –accident causing death | 1 |
| Shipping and navigation -grounding of vessel | 1 |
| Land surveying to settle measurement issues | 3 |
| Employee dispute/ Unfair dismissal | 2 |
| Aircraft accident | 1 |
| Civil accident | 1 |
| Tracking offender under house arrest | 1 |
| Cost distribution | 1 |
| Speeding/ dangerous driving | 1 |

Table 4 – Case Weight presents the raw data for the case assignments. Seven (7) cases were assigned a high weight, 45 a medium weight and 31 a low weight. Figure 8 – Trend lines of Weights assigned presents the weight (High, Medium, and Low) assigned to the GPS evidence in court cases and its variation throughout the years.

**Table 4. Case weight**

| Weight | No. of cases | Percentage |
|---|---|---|
| High | 7 | 8.4 |
| Low | 31 | 37.3 |
| Medium | 45 | 54.2 |

The three data sets for High, Medium and Low don't have linear or polynomial trends. This was confirmed by running an R2 'Goodness-of-fit' test on the High, Medium and Low data [23]. The R2 values lie between 0 and 1, where 0 indicates that Y cannot be predicted by X and 1 indicates that the trend line is the best fit [8]. For this data set, the R2 values for linear trend lines were below 0.6, while the R2 values for the polynomial trend lines were below 0.7. Therefore, the trend lines are shown as four period moving averages. Moving averages are used to reduce random fluctuations and to illuminate underlying trends [23].

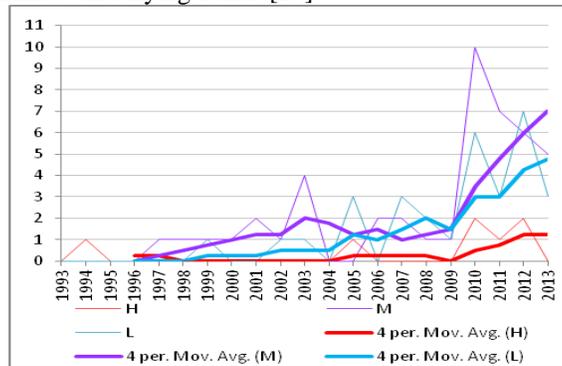

**Figure 8. Trend lines of weights assigned**

The trend line of High weighted evidence shows a slow increase because of its small number of cases. The Medium weighted evidence is the highest in number and has a slightly fluctuating trend. The Low weighted evidence has a higher rate of increase than the High weigh but not as progressive as the Medium with what appears to be slightly small fluctuations.

The relevant 55 criminal cases and the relevant 28 civil cases were examined from an evidence admissibility perspective. Criminal cases showed a higher level of evidence admission with 34 cases compared to 12 admitted civil cases. Civil cases showed a higher level of evidence dismissal with 6 cases while criminal cases only list a single case having dismissed evidence. Table 5 - Evidence admissibility summarizes the data from this analysis. When the criminal and civil admissibility data is considered together, 55% was admitted, 36% not clear and 9 % was dismissed. In most cases, the evidence has been admitted. This is an indication that GPS evidence is being accepted in legal environments. It is interesting that, GPS data inaccuracies were considered in a few cases but the possibility of the deliberately altering GPS records or GPS spoofing was not documented in any case.

**Table 5 - Evidence admissibility**

|  | Criminal cases | | Civil cases | |
| --- | --- | --- | --- | --- |
|  | No. of cases | % | No. of cases | % |
| Evidence Admitted | 34 | 61.8 | 12 | 42.9 |
| Evidence Dismissed | 1 | 1.8 | 6 | 21.4 |
| Not Available | 20 | 36.4 | 10 | 35.7 |

From the above interpretations, it cannot be said that GPS evidence is currently having a high impact in court cases in terms of the weight assignment. However, the large number of cases with Medium weight, which appears to be steadily increasing since 2009, indicates that GPS is increasingly supporting main pieces of evidence in court cases. When criminal and civil cases were taken separately, it was observed that GPS evidence has been present in civil cases years before it appeared in criminal cases. This appears to be due to the use of GPS devices in the maritime industry. The majority of cases found were criminal cases. Also, most of the criminal cases that used GPS evidence are recorded from 2002 onward, while the High weighted criminal cases appear after 2009. This may indicate that GPS evidence is evolving towards having a considerable impact on criminal cases.

### 4.4 Criminal and civil case trends

Figures 9 and 10 present trends in criminal and civil cases, respectively, from 1993 to 2013. The trend line for each figure was drawn as a four period moving average line. A clear increase in criminal cases from 2009 onwards can be seen in the trend line. Extrapolating the 2013 trends indicate there would be approximately 11 new criminal cases and no change in the number of civil cases.

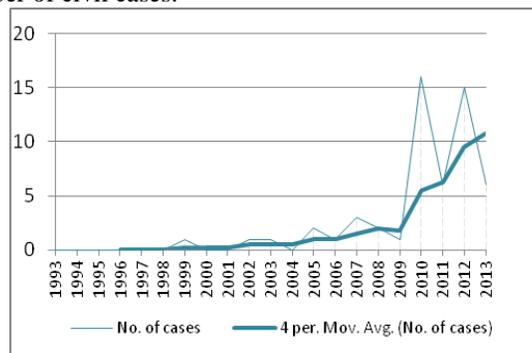

**Figure 9: Trend in criminal cases**

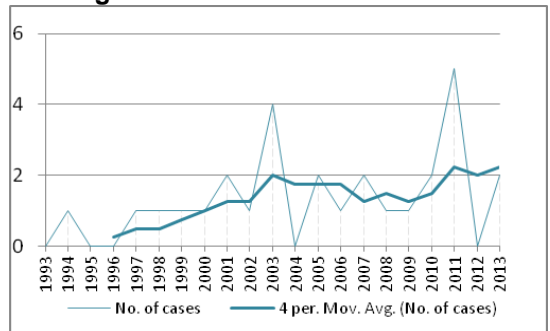

**Figure 10: Trend in civil cases**

### 4.5 Cases according to the decades

Splitting the study's timeline into halves identifies two decades. The first decade is from 01 June 1993 to 01 June 2003 and the second is from 01 June 2003 to 01 June 2013. The first decade had a total of 12 cases. It should be noted that there were two cases in the first half of 2003 that were included in this decade and three cases in the latter half of 2003 that were included in the second decade giving the second decade a total of 71 cases.

The cost associated with the civilian use of GPS technology has been radically reduced between the mid-1980s and the present day [13]. The data collected for this study indirectly reflects the idea that technology cost and utilization are inversely related. The data in the study demonstrates that most of the data was rescored in the second decade as the cost of technology decreased, while interest and utilization increased. Isolating the case types reveals that 53 out of 55 criminal cases and 18 out of 28 civil cases were recorded in the second decade. This translates into 86% of the cases taking place in the second decade while only 14% occurred in the first decade.

### 4.6 Distribution and trend of all cases

Table 6 – Cumulative data provides both the cumulative totals and percentages for the number of cases each year. This shows that approximately 3/4 or 75% of the cases appeared between 2007 and 2013. Table 6 highlights that there is an increase in the number of cases toward the end of the research time-line. Figure 11 presents a line chart of the cases that took place each year. This data does not fit into a linear or polynomial trend line. It was confirmed by doing an $R^2$ 'Goodness-of-fit' test on the data set [23]. Therefore, in the graph below, the trend is shown as a four period moving average line. The graph shows approximately 13 cases by the end of 2013. As shown in Figures 9 and 10, it can be approximated that the 13 cases are comprised of 11 criminal and 2 civil cases.

**Table 6. Cumulative data**

| Year | No. of cases | Cumulative total | Cumulative % |
|---|---|---|---|
| 1993 | 0 | 0 | 0.00% |
| 1994 | 1 | 1 | 1.20% |
| 1995 | 0 | 1 | 1.20% |
| 1996 | 0 | 1 | 1.20% |
| 1997 | 1 | 2 | 2.41% |
| 1998 | 1 | 3 | 3.61% |
| 1999 | 2 | 5 | 6.02% |
| 2000 | 1 | 6 | 7.23% |
| 2001 | 2 | 8 | 9.64% |
| 2002 | 2 | 10 | 12.05% |
| 2003 | 5 | 15 | 18.07% |
| 2004 | 0 | 15 | 18.07% |
| 2005 | 4 | 19 | 22.89% |
| 2006 | 2 | 21 | 25.30% |
| 2007 | 5 | 26 | 31.33% |
| 2008 | 3 | 29 | 34.94% |
| 2009 | 2 | 31 | 37.35% |
| 2010 | 18 | 49 | 59.04% |
| 2011 | 11 | 60 | 72.29% |
| 2012 | 15 | 75 | 90.36% |
| 2013 | 8 | 83 | 100.00% |

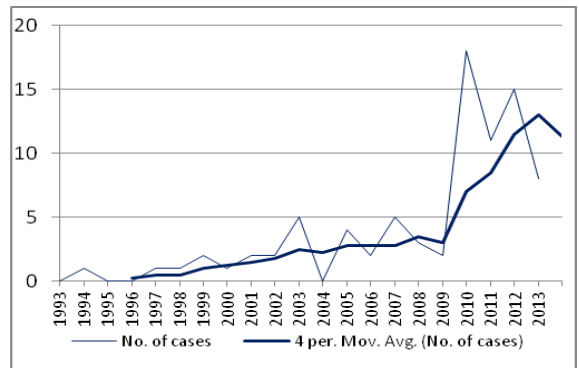
**Figure 11. Case Trends**

## 5. Conclusions and future work

The increased amalgamation of GPS enabled technology into highly networked societies raises the need to understand the legal impact. This case study presents initial research exploring the quantification of GPS data in court cases, analysis of the types of crimes in conjunction with modes of transportation along with impact and trend identification.

Quantification of the data reveals that GPS evidence is being introduced in a variety of criminal and civil cases and that this evidence is increasingly impacting digital evidence in court. The ratio of the number of cases, between the first and second decades, is 12:71 indicating that the number of cases have increased dramatically since June 1, 2003. It is worth noting that 75% of the cases were recorded from 2007 onwards. Overall, the same trend is true when considering criminal and civil cases separately. The number of criminal cases involving GPS evidence shows a steady increase from 2009 onwards. The results from this study reveal that the number of cases involving GPS evidence has dramatically increased during the last decade and has an increasing trend, while playing a considerable role in court case verdicts.

Analysis of the data illustrates that it is possible to correlate the types of criminal cases with specific modes of transportation. Devices used on land appeared mostly in criminal cases, while devices used in water based transportation systems appeared mostly in civil cases. The data also indicates that there is a possible trend in criminal drug cases to utilize GPS data. It also reveals a possible trend in civil cases to utilize GPS data in ship collisions.

The weights high, medium and low were associated with terms to categorize qualitative data in order to gage the criticality of the evidence. The application of classifications to the data set reveals that out of the 83 relevant cases 62.6% of the cases are 'Medium' or 'High' weighted and that the highest individual rating was for 'Medium' weighted cases. This indicates that the main piece of evidence was supported by GPS evidence in these cases. It is also interesting that the data analysis of the four period moving averages for 'Medium' and 'High' trends show an increase from 2009 onwards. As GPS technology continues to become progressively integrated into other devices and increasingly affordable, it will be interesting to see if GPS data in the future has a higher impact on legal cases.

The overall analysis of the data supports the hypotheses that the use of GPS evidence in court proceedings has increased during the past decade and appears to be playing an important role in court cases. It is interesting that there was a lack of information readily available on the experts or the detailed findings of their analysis in the databases used for this research. A more detailed examination of the technical reports and the different methodologies used to analyze GPS evidence by different experts may reveal additional insights into usage patterns and data storage locations. Therefore, future work should acquire and conduct further research into the expert reports involving GPS evidence, generated for individual court cases.

Furthermore, research can be conducted on other types of devices such as solid state drives, tablets and cloud storage applications to ascertain the impact of the data extracted from these devises and software solutions are having in legal environments. From a software engineering perspective, research should also examine the reliability of GPS data extraction software solutions in terms of verifiability. These research ideas could be extended to examine court case reports from other countries that have similar, more or less mature technology markets. This could possibly identify international trends.

Future work should examine individual modes of transportation to investigate GPS technology corroboration with other potential location identification technologies such as Wi-Fi, cell towers and social media interaction. The research should examine the security and privacy issues in conjunction with opportunities to impact cyber physical systems.